\documentclass[aps,pre,twocolumn,superscriptaddress]{revtex4-2}
\usepackage{natbib}
\usepackage{mathrsfs}
\usepackage{graphicx}
\usepackage{bm}
\usepackage{threeparttable}
\usepackage{amsmath,amssymb}
\usepackage{color}
\usepackage{epstopdf}
\usepackage[hidelinks]{hyperref}
\usepackage{changes}

\definechangesauthor[color=red]{diver}

\begin{document}

\title{Intrinsic pressure as a convenient  mechanical framework for dry active matter}

\author{Zihao Sun}
\thanks{These authors contributed equally to this work.}
\affiliation{Beijing National Laboratory for Condensed Matter Physics and Laboratory of Soft Matter Physics, Institute of Physics, Chinese Academy of Sciences, Beijing 100190, China}
\affiliation{School of Physical Sciences, University of Chinese Academy of Sciences, Beijing 100049, China}
\author{Longfei Li}
\thanks{These authors contributed equally to this work.}
\affiliation{Beijing National Laboratory for Condensed Matter Physics and Laboratory of Soft Matter Physics, Institute of Physics, Chinese Academy of Sciences, Beijing 100190, China}
\author{Chuyun Wang}
\thanks{These authors contributed equally to this work.}
\affiliation{Postgraduate training base Alliance, Wenzhou Medical University, Wenzhou,
Zhejiang 325035, China}
\author{Jing Wang}
\affiliation{Wenzhou Institute, University of Chinese Academy of Sciences, Wenzhou, Zhejiang 325001, China}
\author{Huaicheng Chen}
\affiliation{Wenzhou Institute, University of Chinese Academy of Sciences, Wenzhou, Zhejiang 325001, China}
\author{Gao Wang}
\affiliation{Wenzhou Institute, University of Chinese Academy of Sciences, Wenzhou, Zhejiang 325001, China}
\author{Liyu Liu}
\email[lyliu@cqu.edu.cn]{}
\affiliation{Human Phenome Institute, Fudan University, Shanghai, 201203, China}
\affiliation{Wenzhou Institute, University of Chinese Academy of Sciences, Wenzhou, Zhejiang 325001, China}
\author{Fangfu Ye}
\email[fye@iphy.ac.cn]{}
\affiliation{Beijing National Laboratory for Condensed Matter Physics and Laboratory of Soft Matter Physics, Institute of Physics, Chinese Academy of Sciences, Beijing 100190, China}
\affiliation{School of Physical Sciences, University of Chinese Academy of Sciences, Beijing 100049, China}
\affiliation{Wenzhou Institute, University of Chinese Academy of Sciences, Wenzhou, Zhejiang 325001, China}
\affiliation{Oujiang Laboratory (Zhejiang Lab for Regenerative Medicine, Vision and Brain Health), Wenzhou, Zhejiang 325000, China}
\author{Mingcheng Yang}
\email[mcyang@iphy.ac.cn]{}
\affiliation{Beijing National Laboratory for Condensed Matter Physics and Laboratory of Soft Matter Physics,
Institute of Physics, Chinese Academy of Sciences, Beijing 100190, China}%
\affiliation{School of Physical Sciences, University of Chinese Academy of Sciences, Beijing 100049, China}

\begin{abstract}
The identification of local pressure in active matter systems remains a subject of considerable debate. Through theoretical calculations and extensive simulations of various active systems, we demonstrate that intrinsic pressure (defined in the same way as in passive systems) is an ideal candidate for local pressure of dry active matter, while the self-propelling forces on the active particles are considered as effective external forces originating from the environment. Such a framework is universal and especially convenient for analyzing mechanics of dry active systems, and it recovers the conventional scenario of mechanical equilibrium well-known in passive systems. Thus, our work is of fundamental importance to further explore mechanics and thermodynamics of complex active systems.
\end{abstract}

\date{\today}
\pacs {}
\maketitle

\section{Introduction}

Active matter is composed of self-propelling units, ranging from bacterial colonies, artificial active colloids to bird flocking~\cite{Ramaswamy2010arcmp,Bechinger2016rmp}. Due to their inherently nonequilibrium nature, active matter systems often exhibit intriguing phenomena, such as motility-induced phase separation (MIPS)~\cite{Tailleur2008prl,Cates2015arcmp} and large-scale collective motion~\cite{Vicsek1995prl,Vicsek2012pr,Bricard2013nature,Baconnier2022np}, which are absent in passive systems. Despite significant progress over the past two decades, there is still no consensus on the understanding of some fundamental physical concepts of active matter. A prominent example is what is the local pressure in active systems.

Currently, the dominant view is that active pressure is considered as the local pressure of active matter. The active pressure, $P_\text{a}$, contains three contributions~\cite{Yang2014sm,Takatori2014prl,Mallory2014pre,Winkler2015sm,Solon2015prl,Solon2015np,
Bialke2015prl,Ezhilan2015jfm,Ginot2015prx,Takatori2016nc,Falasco2016njp,Marconi2016sm,Takatori2017prf,
Levis2017sm,Patch2017pre,Marconi2017jcp,Yan2018njp,Ginot2018njp,Wittmann2019jcp,Das2019sr,Soker2021prl,
Auschra2021pre,Wysocki2022njp,Pellicciotta2023nc,Cameron2023pre}: the swim pressure $P_\text{s}$ generated by the persistent self-propelling force of active particles, the kinetic pressure $P_\text{k}$ due to particle motion, and  the interacting pressure $P_\text{c}$ from direct interparticle interactions, i.e., $P_\text{a} = P_\text{s} + P_\text{k} + P_\text{c}$. In this context, the self-propelling force experienced by the active particle is regarded as an internal force. Active pressure provides important insights into various aspects of active systems, such as the phase behavior of active particles~\cite{Takatori2014prl,Solon2015prl,Takatori2015pre,Levis2017sm,Solon2018njp,
Digregorio2018prl,Speck2021pre,Mallory2021pre}, and the motion~\cite{Yan2015jfm,Razin2017pre,
Duzgun2018pre,Li2022jcp}, deformation~\cite{Nikola2016prl,Paoluzzi2016sr,Li2019prl,
Wang2019jcp,Vutukuri2020nature,Quillen2020pre} and effective interactions~\cite{Ray2014pre,Ni2015prl,Liu2020prl,Paul2022prl,Li2022sm} of passive objects immersed in active baths.

However, employing active pressure in the analysis of mechanical equilibrium in active matter often leads to counterintuitive results. For instance, in the presence of boundary walls, the bulk active pressure generally is not equal to the mechanical pressure on the wall exerted by active particles (the wall pressure $P_\text{w}$, depending on the wall details), except for the special case of spherical (torque-free) active Brownian particles (ABPs) confined by planar walls~\cite{Takatori2014prl,Mallory2014pre,Fily2014sm,Wysocki2015pre,Yan2015jfm,Solon2015prl,
Solon2015np,Nikola2016prl,Fily2017jpa,Junot2017prl,Jamali2018sm,Wittmann2019jcp,Sandoval2020pre,
Row2020pre,Xie2022prl,Sandoval2023sm,Liu2023jpcm}. This implies a general absence of a state function for active pressure~\cite{Solon2015np,Wysocki2015pre,Fily2017jpa,Junot2017prl,Liu2023jpcm}.
Additionally, in a system with a spatially inhomogeneous activity, the bulk active pressures in the low- and high-activity regions are also unequal~\cite{Solon2015np,Row2020pre}, apparently implying that the system would experience a net force. While the pressure imbalances may be reconciled by invoking active source/sink~\cite{Fily2017jpa} or the flux of polar order~\cite{Wysocki2022njp}, such treatments introduce extra complexity and are not intuitive. More surprisingly, the surface tension of phase-separated ABPs, calculated based on the active pressure, is significantly negative~\cite{Bialke2015prl}, seemingly contradicting with the stable MIPS. The counterintuitive results and related complexities encountered in the $P_\text{a}$-based mechanical equilibrium prompt consideration of the local pressure of active matter from an alternative perspective. 

Several recent studies have proposed that the local pressure in active systems should only consist of kinetic and interacting pressures, i.e. $P_\text{i} = P_\text{k} + P_\text{c}$, the same as the traditional definition of pressure in passive systems (termed {\it intrinsic} pressure in this article); while the self-propelling force is treated as an external force arising from surrounding environment~\cite{Yan2015sm,Speck2016pre,Rodenburg2017sm,Steffenoni2017pre,Paliwal2018njp,
Epstein2019jcp,Omar2020pre,Speck2020sm,Lauersdorf2021sm,Omar2023pnas}. In this framework, the possible mismatch of pressure in different regions, whether in the systems with boundary walls or inhomogeneous activities, could be naturally balanced by the effective external force emerging from the self-propelling forces near interface (i.e., polarization force), without needing extra quantities. This restores the conventional scenario of mechanical equilibrium. Moreover, applying the intrinsic pressure to the phase-separated ABPs can lead to a positive surface tension~\cite{li2023surface}. The intrinsic pressure looks a promising candidate for the local pressure of active matter, but its validity has only been tested by simulations for the minimal ABP models, in which spherical particles couple each other via torque-free potential interactions ~\cite{Yan2015sm,Speck2016pre,Lee2017sm,Rodenburg2017sm,Paliwal2018njp,Omar2020pre,Lauersdorf2021sm,Omar2023pnas}. However, the particle couplings in real active matter systems are quite diverse and complex, including alignment, quorum sensing, communication, nonreciprocal interactions and so on. Therefore, in order to verify the universality of this theoretical framework, it is necessary to examine the $P_\text{i}$-based mechanical equilibrium across a broader range of active systems with different types of interactions.

In this work, we combine theoretical analyses and computer simulations to explore the steady-state force balance in various dry active systems, based on the framework of intrinsic pressure. These systems include those with boundary walls, nonuniform activity, as well as diverse types of interactions such as alignment and communication. We show that the conventional mechanical equilibrium condition holds across all systems, wherein the existing pressure imbalance is exactly compensated by the external polarization force. Thus, our results strongly suggest that it is more convenient and natural to adopt the intrinsic pressure as the local pressure of dry active matter. 

\section{Theory and simulations}

We consider a two-dimensional overdamped dry active system composed of generalized ABPs, where the self-propelling velocities of active particles (both magnitudes and orientations) may depend on the surrounding environment due to sensing or communication. The dynamics of particle $i$, with position $\mathbf{r}_i = (x_i,y_i)$ and orientation $\mathbf{e}_i = (\cos\theta_i,\sin\theta_i)$, is generally described by
\begin{align}\label{eq1}
	\dot{\mathbf{r}}_i &= v_0 \mathbf{e}_i + \frac{1}{\gamma_\text{t}} (\sum_{i \neq j} \mathbf{F}_{ij} + \mathbf{F}^\text{ext}_i)+ \sqrt{2D_\text{t}}{\boldsymbol \eta}_i\nonumber\\
	\dot{\theta}_i &= \frac{\Gamma_0}{\gamma_\text{r}} + \frac{1}{\gamma_\text{r}} (\sum_{i \neq j} \Gamma_{ij} + \Gamma^\text{ext}_i) + \sqrt{2D_\text{r}}\xi_i,
\end{align}
with $v_0$ the self-propelling velocity and $\Gamma_0$ the self-rotating torque of active particle. Here, $D_\text{t} = k_\text{B}T/ \gamma_\text{t}$ represents the translational diffusion coefficient with the translational friction coefficient $\gamma_\text{t}$ and thermal energy $k_\text{B}T$; $\mathbf{F}_{ij}$ and $\mathbf{F}^\text{ext}_i$ separately denote the interparticle interaction and the external force; and ${\boldsymbol \eta}_i$ refers to Gaussian-distributed white noises of zero mean and unit variance. Similarly, $D_\text{r} = k_\text{B}T / \gamma_\text{r}$, $\gamma_\text{r}$, $\Gamma_{ij}$, $\Gamma^\text{ext}_i$ and $\xi_i$ are the corresponding quantities in the rotational degrees of freedom. 

From Eq.~(\ref{eq1}), one can derive the Smoluchowski equation characterizing the evolution of noise-averaged probability distribution function $\psi(\mathbf{r},\theta,t) = \langle \hat{\psi}(\mathbf{r},\theta,t)\rangle = \langle \sum_i \delta(\mathbf{r} - \mathbf{r}_i(t))\delta(\theta - \theta_i(t))\rangle$~\cite{Solon2015prl,Solon2015np,Paliwal2018njp,Wysocki2022njp}
\begin{align}\label{eq2}
	\partial_t\psi &= -\frac{1}{\gamma_\text{t}}\mathbf{\nabla}\cdot \left[ \gamma_\text{t} v_0\mathbf{e}\psi  + \mathbf{f}^\text{ext}\psi + \int d\mathbf{r}' \int d\theta' \mathbf{f} \langle \hat{\psi}'\hat{\psi} \rangle\right]\nonumber \\
	& - \frac{1}{\gamma_\text{r}} \partial_{\theta}\left[\Gamma_0\psi + \Gamma^\text{ext}\psi + \int d\mathbf{r}'\int d\theta' \Gamma\langle \hat{\psi}'\hat{\psi} \rangle \right] \nonumber\\
	& + D_\text{t}\mathbf{\nabla}^2\psi +  D_\text{r} \partial^2_{\theta}\psi,
\end{align}
where the interparticle force and torque are rewritten as $\mathbf{f}(\mathbf{r'},\theta',\mathbf{r},\theta)$ and $\Gamma(\mathbf{r'},\theta',\mathbf{r},\theta)$, respectively, and the external counterparts are $\mathbf{f}^\text{ext}(\mathbf{r},\theta)$ and $\Gamma^\text{ext}(\mathbf{r},\theta)$. For notational brevity in the integrals, we denote $\hat{\psi}' \equiv \hat{\psi}(\mathbf{r}',\theta',t)$. Assuming translational invariance along the $y$-axis, integrating Eq.~(\ref{eq2}) over $\theta$ in the steady state yields  $dJ_\text{n} / dx = 0$, with $J_\text{n}$ the particle flux along the $x$-direction reading
\begin{align}\label{eq3}
	\gamma_\text{t} J_\text{n} &= f_\text{m}  + \int d\theta f^\text{ext}_x \psi + I_1  - k_\text{B}T \frac{d  n}{d x},
\end{align}
with
\begin{align}
	I_1(x) &\equiv \int d\theta \int d\mathbf{r}' \int d\theta' f_x(\mathbf{r},\theta,\mathbf{r}',\theta') \langle\hat{\psi}'\hat{\psi}\rangle\nonumber.
\end{align}
Here, $n(x) = \int d\theta \psi(x,\theta)$ and $f_\text{m}(x) = \int d\theta \gamma_\text{t} v_0 \cos\theta \psi(x,\theta)$ represent the number density and polarization force density, respectively, and $I_1(x)$ accounts for the contribution from the interparticle interactions. Note that the kinetic pressure in the overdamped limit is $P_\text{k}(x) = n(x) k_\text{B} T$~\cite{Speck2016pre}, and the integral of $I_1(x)$ is exactly the interacting pressure $P_\text{c}(x) = \int_x^{\infty} I_1(x') dx'$~\cite{Solon2015prl,Wysocki2022njp}. Thus, Eq.~(\ref{eq3}) just corresponds to local force balance 
\begin{equation} \label{eq4}
	\frac{d P_\text{i}}{d x}  = f_\text{m} + \int d\theta f^\text{ext}_x \psi - \gamma_\text{t} J_\text{n},
\end{equation}
with the intrinsic pressure $P_\text{i} = P_\text{k} + P_\text{c}$. Clearly, the local pressure is $P_\text{i}$, and the polarization force density $f_\text{m}$, acting as an external force, participates in the force balance of the system. The second and third terms of Eq.~(\ref{eq4}) correspond to other external forces (such as gravity and wall interactions) and friction, respectively. As stated in the {\it Introduction}, mechanical equilibrium, in general, cannot be intuitively established based on active pressure $P_\text{a}$ in the presence of the boundary wall and spatially inhomogeneous activity. However, as we shall see, this issue can be naturally addressed when using intrinsic pressure $P_\text{i}$.

\subsection{Systems with wall boundary}
\begin{figure}[t]
	\centering 
	\includegraphics[width = .48\textwidth]{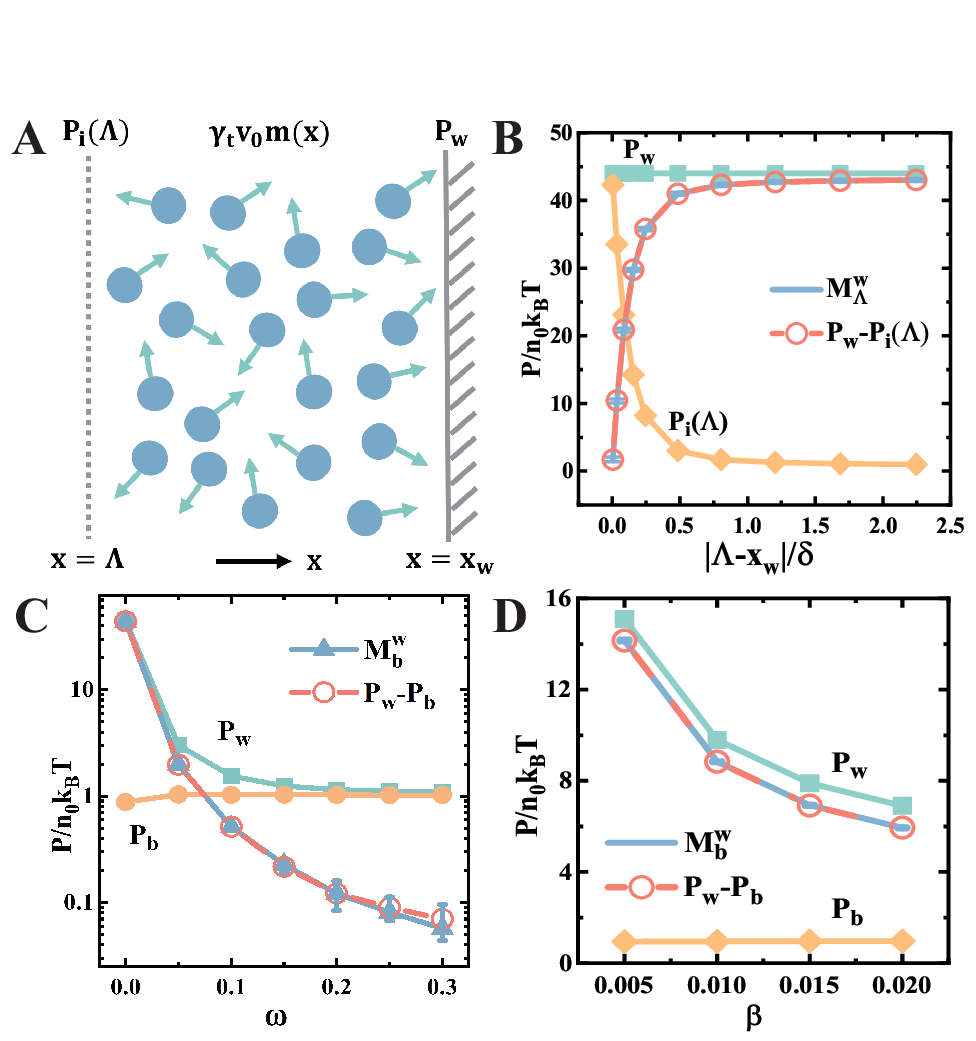}
	\caption{(a) Schematic diagram of overdamped ABPs confined by boundary walls. (b) Pressure profile of torque-free ideal ABPs near a hard wall. Here, the horizontal coordinate is reduced by a characteristic length $\delta = \sqrt{D_\text{t}/D_\text{r}}$. The verification of Eq.~(\ref{eq6}) for ideal ABPs (c) with self-propelling angular velocity $\omega = \Gamma_0/\gamma_\text{r}$ or (d) subjected to an external torque $\Gamma_i^\text{ext} = \beta F^\text{w}_i \sin2\theta_i$ due to the wall, with $\beta$ being the interaction strength. The force $F^\text{w}_i$ exerted by the wall on particle $i$ is given by $F^\text{w}_i = -dU/dx$, with a Weeks-Chandler-Andersen (WCA) potential $U = 4\epsilon \left[(\sigma/|x-x_\text{w}|)^{12} - (\sigma/|x - x_\text{w}|)^6\right] + \epsilon$ if the particle-wall distance $|x-x_\text{w}| < 2^{1/6}\sigma$ ($\sigma = 1$, $\epsilon = 1$). For all plots, we use $\gamma_\text{t} = \gamma_\text{r} = 100$, $v_0 = 0.1$ and the global number density $n_0 = 0.1$.}
	\label{fig1}
\end{figure}

Consider overdamped ABPs confined by a vertical wall locating at $x = x_\text{w}$, see Fig.~\ref{fig1}(a). The wall pressure $P_\text{w}$ exerted by ABPs can be readily expressed as:
\begin{equation}\label{eq5}
	P_\text{w} = -\int_{\Lambda}^{x_\text{w}} dx \int d\theta f^\text{w}_x(x,\theta) \psi(x,\theta),
\end{equation}
where the wall-particle interaction $f_x^\text{w}(x,\theta)$ vanishes at $x = \Lambda$ and is assumed to be infinitely large at $x = x_\text{w}$ such that $\psi(x_\text{w},\theta) = 0$. The impenetrable wall also implies $J_\text{n}(x) = 0$ and $P_\text{i}(x_\text{w}) = 0$. Therefore, in the absence of other external forces, integrating Eq.~(\ref{eq4}) from $x = \Lambda$ to $x = x_\text{w}$, we obtain the force balance formula:
\begin{equation}\label{eq6}
	P_\text{w} - P_\text{i}(\Lambda) = \int_{\Lambda}^{x_\text{w}} f_\text{m} dx = M_{\Lambda}^\text{w},
\end{equation} 
with $M_\Lambda^\text{w}$ being defined as the integral of polarization force density. Notably, the local pressure $P_\text{i}(\Lambda)$ automatically becomes the bulk pressure $P_\text{b}$ of the active system when $\Lambda$ locates inside the bulk region. Equation~(\ref{eq6}) clearly indicates that the polarization force is the root cause of the discrepancy between the wall pressure and local pressure. To demonstrate the universality of the force balance in Eq.~(\ref{eq6}), we firstly investigate the case of the minimal non-interacting ABPs confined by a hard wall through simulations, see Fig.~\ref{fig1}(b), and then turn to the active systems subjected to either the self-propelled torque or externally applied torque, as illustrated in Fig.~\ref{fig1}(c) and \ref{fig1}(d). The results show that the difference between the local pressure $P_\text{i}(\Lambda)$ and wall pressure $P_\text{w}$ consistently aligns with the integral of polarization force density $M_\Lambda^\text{w}$, as expected. Note that all data presented in this article are obtained from Brownian dynamics simulations with a fixed thermal energy $k_\text{B}T = 1$.

\subsection{System with nonuniform activity}

\begin{figure}[tbh]
	\centering 
	\includegraphics[width = .48\textwidth]{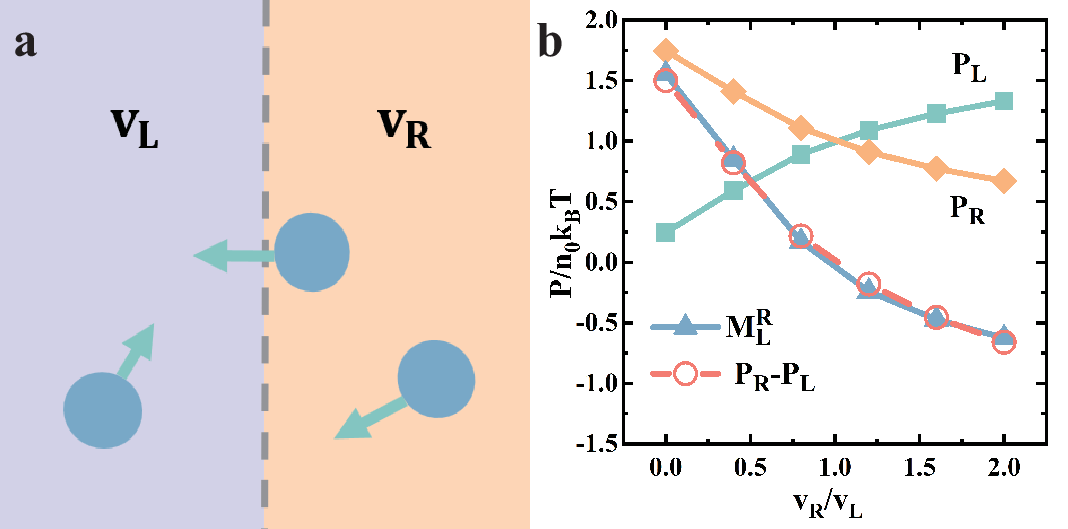}
	\caption{(a) Sketch of active system with piecewise activity, where the propulsion velocity of a particle is $v_\text{L}$ in the left region and $v_\text{R}$ in the right region. (b) The bulk pressures, the pressure difference, and the integral of polarization force density as functions of the ratio $v_\text{R}/v_\text{L}$ for ideal overdamped ABPs, with $n_0 = 0.1$, $\gamma_\text{t} = \gamma_\text{r} =100$ and $v_\text{L} = 0.1$. }
\label{fig2}
\end{figure}
Apart from systems with the walls, the intrinsic pressure can also naturally maintain mechanical equilibrium in systems with inhomogeneous activity. Consider an ideal overdamped ABP system with a piecewise activity pattern, where the self-propelling velocity of particles is set to $v_\text{L}$ when they are in the left region and to $v_\text{R}$ otherwise, as sketched in Fig.~\ref{fig2}(a). In flux-free steady state, the spatial variations in activity lead to variations in concentration, $n \propto 1/v$~\cite{Cates2015arcmp,arlt2019dynamics,Row2020pre}. However, the swim pressure satisfies $P_\text{s} \propto n v^2$, so that the bulk active pressures $P_\text{a}$ on the two sides are unequal~\cite{Solon2015np,Row2020pre,Soker2021prl,Wysocki2022njp}. Instead, as shown in Fig.~\ref{fig2}(b), the disparity in activity results in significant polarization near the interface, which compensates for the difference in the intrinsic pressure between the left and right regions. This force balance condition is straightforwardly predicted by integrating Eq.~(\ref{eq4}), 
\begin{equation}\label{eq7}
\left. P_\text{i} \right|_\text{R}-\left. P_\text{i} \right|_\text{L} = \int_\text{L}^\text{R} f_\text{m} dx = M_\text{L}^\text{R},
\end{equation}
where $L$ and $R$ are taken deep into the corresponding bulk phases.

\subsection{Systems with communicating interaction}

Although we numerically demonstrate the validity of Eq.~(\ref{eq7}) only in the ideal overdamped ABP systems, it remains applicable in the presence of any interparticle interactions, including steric and communicational cases. To highlight this point, we shift our focus to the ABPs with communication interactions~\cite{borra2021optimal,lavergne2019group,karamouzas2014universal}. In this system, particles can exert non-contact {\it effective} forces or torques on each other by detecting the positions and velocities of their neighbors. We here explore three typical communication systems and evaluate the validity of intrinsic pressure acting as the local pressure. The simulations are performed in a square box with periodic boundary conditions, in which the box is divided into two regions $L$ and $R$, characterized by different parameters. Additionally, there is a short-range repulsive interaction between the particles, which is modeled using the WCA potential. For these three communication systems, we numerically measure the bulk intrinsic pressures $P_\text{L}$ and $P_\text{R}$ in regions $L$ and $R$, respectively, along with the corresponding integral of the polarization force. In contrast, the mechanical equilibrium based on active pressure in such systems would be elusive. 

\begin{figure}[tbh]
\centering 
\includegraphics[width = .48\textwidth]{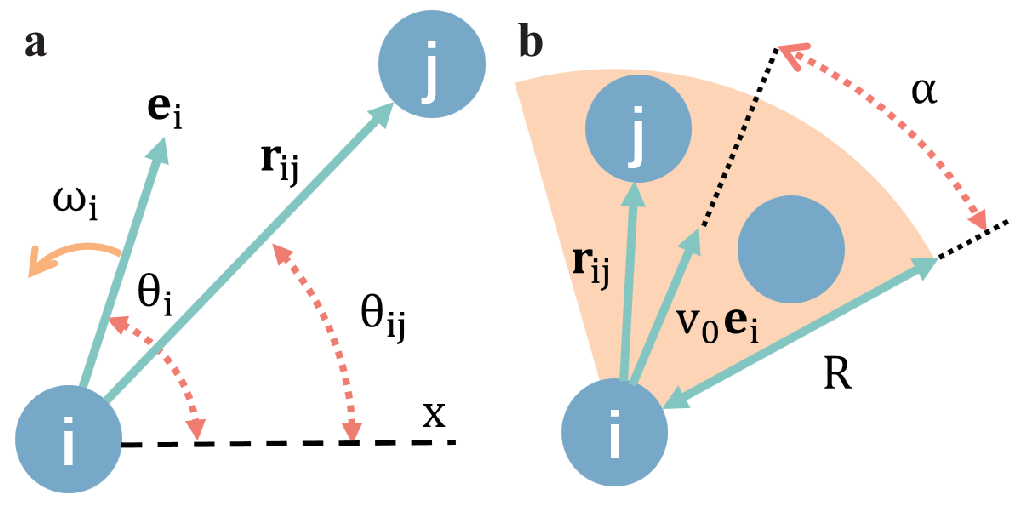}
\caption{Communication interactions. (a) Self-avoiding model: Particle $i$ rotates its propulsion direction ${\mathbf e}_i$ counterclockwise if $\theta_i \ge \theta_{ij}$ (clockwise otherwise), to avoid collisions with particle $j$. (b) Visual perception model: Particle $i$ actively adjusts its velocity $v_0$  based on a perception function that depends on the degree of crowding within a vision cone of half-width angle $\alpha$ and radius $R$.}
\label{fig3}
\end{figure}

\emph{Rotational self-avoiding particles}.\textemdash In groups of organisms, collisions are often avoided through complex, coordinated movements. To simulate this phenomenon, we employ a rotational self-avoiding ABP model~\cite{borra2021optimal}. In this model, upon reaching sensing distance $\sigma_\text{s}$, active particles start to adjust their direction to avoid collisions, as sketched in Fig.~\ref{fig3}(a). The angular speed of steering, $\omega_i$, is determined by the distance between particles and is given by:
\begin{equation}\label{eq8}
\omega_i = \sum_j \mathrm{sgn}(\theta_i - \theta_{ij}) \, \omega_0 \left( \frac{\sigma_\text{s}}{|\mathbf{r}_{ij}|} \right)^2,
\end{equation}
for $|\mathbf{r}_{ij}| < \sigma_\text{s}$. Here, $\mathrm{sgn}(x)$ represents sign function, and $\theta_{ij}$ is the angle between $\mathbf{r}_{ij}$ and the $x$-axis. Besides, $\omega_0$ quantifies the particles' ability to avoid collisions. 

\emph{Visual perception particles}.\textemdash Visual perception, which occurs frequently in living systems, is a prominent class of communication interactions. Here, we draw on the visual perception model proposed by Bechinger \textit{et al.} ~\cite{lavergne2019group}, see Fig.~\ref{fig3}(b). The perception function is given by:
\begin{equation}\label{eq9}
P_i = \sum_{j\ \in V_i^\alpha} \frac{1}{r_{ij}},
\end{equation}
where $V_i^\alpha$ represents the visual cone. When the perception is below a threshold $P^\star$, the active particles maintain their self-propelling velocity $v_0$. Otherwise, the self-propelling velocity is reduced to $v_0 \mathrm{exp}[-(P - P^\star)R]$.
The threshold $P^\star$ depends on the global number density $n_0$ and perception radius $R$, $P^{\star} = \alpha n_0R/\pi$. In the cases where $\alpha < \pi$, the asymmetric vision cones between particles give rise to a nonreciprocal inter-individual communication effect.

\emph{Humanoid interactive particles}.\textemdash As a typical active system, the crowded pedestrian often emerges fascinating collective motion possibly arising from their effective interactions related to expected collision times, as introduced by Ref.~\cite{karamouzas2014universal}. Specifically, the effective interaction between particles is
\begin{equation}\label{eq10}
F_\text{p} = -\nabla_{\mathbf{r}} \left( \frac{k}{\tau^2} e^{-\frac{\tau}{\tau_0}}\right),
\end{equation}
where $k$ represents the interaction strength, $\tau$ denotes the expected collision time determined by linearly extrapolating pedestrians’ trajectories, based on current relative velocities $\mathbf{v}$ and displacements $\mathbf{r}$ (specific calculation rules can be found in Ref.~\cite{karamouzas2014universal}), and $\tau_0$ indicates the maximum interaction range. It is important to note that this pairwise force, $F_\text{p}$, is not a direct interaction between the particles but a response to the communication between them. From this perspective, $F_\text{p}$ can only be contributed by the environment (such as the substrate) and modify the self-propelling force, so that it should be understood as an external force. Thus, it can be expected that a nonzero external force, $\Pi_\text{L}^\text{R} = \int_\text{L}^\text{R} F_\text{p}(x) dx$, will emerge near the interface between subregions $L$ and $R$, as the case of the polarization force.

\begin{figure*}[!bht]
\centering 
\includegraphics[width = .9\textwidth]{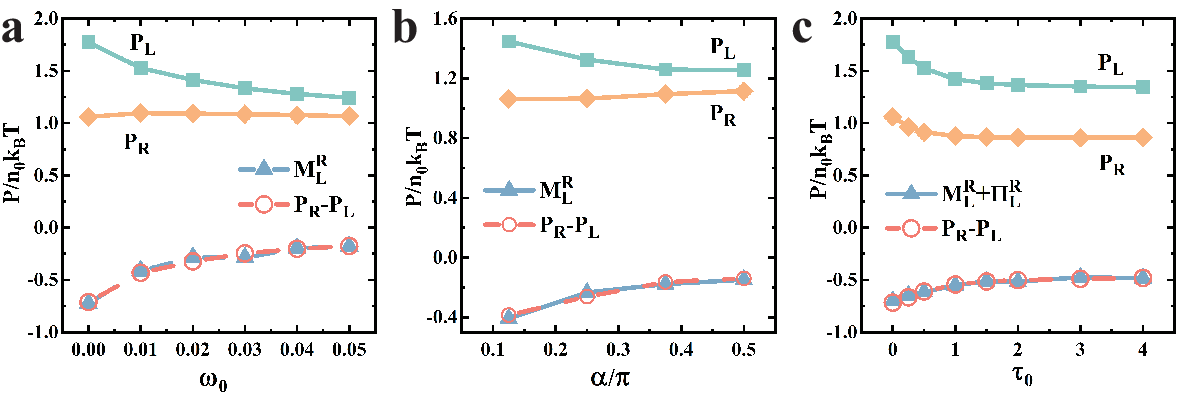}
\caption{Intrinsic pressures, pressure difference and effective external force in communication systems with a piecewise activity landscape, $v_\text{L} = 0.1$ and $v_\text{R} = 0.2$ (subscripts $L$ and $R$ refer to the left and right half regions). Here, the horizontal coordinates separately correspond to coupling parameters in three different  communication models. (a) Rotational self-avoiding particles with $\sigma_\text{s} = 3\sigma$. (b) Visual perception particles with $R = 3\sigma$. (c) Humanoid interactive particles with $k = 3$. In all cases, we maintain $\sigma = 1$, $\phi = n_0\pi\sigma^2 = 0.1$, and $\gamma_\text{t} = 30$, $\gamma_\text{r} = 10$.} 
\label{fig4}
\end{figure*}

Communication interactions can alter both the direction and amplitude of the propulsion velocity of active units, thus modifying the self-propelling force on them. However, these effects have no influence on the direct interactions among the active particles. Consequently, there is nothing new beyond the intrinsic pressure and the contribution of external force. In terms of the three typical communication rules, we perform simulations to quantify the intrinsic pressure and the emerged external force, as plotted in Fig.~\ref{fig4}. The simulation results unambiguously show that the traditional mechanical equilibrium condition is still valid for the communication systems based on the framework of intrinsic pressure, which consolidates our conclusions.

\section{Discussions}

The validity of the $P_\text{i}$-based framework in mechanical equilibrium of dry active matter has been evaluated by simulations. We now justify the point that the self-propelling force is regarded as an effective external force. Active systems are often described as being free from external forces, but this statement is correct only when considering the active system as a whole. In fact, an active matter system at least comprises active particles and their surrounding environment (such as solvent or substrate), thus amounting to a two-component system. When the discussion is limited to the mechanics of active particles themselves, the self-propelling forces, exerted by the surrounding environment, are definitely classified as external forces. Even the `dry' active matter system, as the case of the present work, should also be understood as an effective binary system, in which the environment is not explicitly taken into account but contributes friction, stochastic force and self-propelling (body) force on the active particles. In this sense, the local pressure of the active particles naturally adheres to the traditional definition in passive systems, excluding the contribution of the self-propelling force. 
 
Finally, we compare the $P_\text{i}$-based framework with the widely-used one based on the active pressure $P_\text{a}$. For simplicity, the magnitude of the self-propelled velocity is assumed to be orientation-independent. The conservation equation of polar order $m(x,t) = \int {d \theta} \cos \theta \psi(x,\theta,t)$ can be derived through multiplying Eq.~(\ref{eq2}) by $\cos\theta$ and then integrating over $\theta$ :
\begin{align}\label{eq11}
\frac{\partial m}{\partial t} = &-\frac{\partial J_\text{m}}{\partial x} - D_\text{r}m \\
&-\frac{1}{\gamma_\text{r}}\int d\theta \sin\theta \left[(\Gamma_0 + \Gamma^\text{ext})\psi + \int d\mathbf{r}'\int d\theta' \Gamma \langle\hat{\psi}'\hat{\psi}\rangle\right], \nonumber
\end{align}
with the flux of polar order (microscopic expression $J _\text{m}(\mathbf{r},t) \equiv \langle \sum_i \dot{\mathbf{r}}_i(t)\mathbf{e}_i(t) \delta(\mathbf{r} - \mathbf{r}_i(t))\rangle$)
\begin{align}
\gamma_\text{t} J_\text{m} &= \gamma_\text{t} v_0\left[\frac{n}{2} + Q\right] - k_\text{B}T\frac{\partial m}{\partial x} + \int d\theta f^\text{ext}_x \cos\theta\psi + I_2,\nonumber
\end{align}
where $I_2(x,t) \equiv \int d\theta \cos \theta \int d\mathbf{r}' \int d\theta' f_x(\mathbf{r},\theta,\mathbf{r}',\theta') \langle\hat{\psi}'\hat{\psi}\rangle$, and $Q(x,t) = \int d\theta \cos2\theta \psi(x,\theta,t)/2$ is the nematic order. The first divergence term on the right-hand side of Eq.~(\ref{eq11}) represents the  contribution from the flux of polar order, while the other two terms characterize sources/sinks  of polar order originating from angular diffusion and torques, respectively. 

In terms of the flux of polar order, the swim pressure can be expressed as $\gamma_\text{t} v_0 J_\text{m}(x,t)/D_\text{r}$~\cite{Fily2017jpa,Paliwal2018njp,Das2019sr,Speck2020sm,Omar2020pre,Wysocki2022njp}. In this way, plugging the steady-state Eq.~(\ref{eq11}) (with $\partial_t m = 0$) into Eq.~(\ref{eq4}) and integrating by parts leads to the force balance for the active pressure, $P_\text{a} = P_\text{i} + P_\text{s}$,
\begin{align}\label{eq13}
\frac{d P_\text{a}}{d x}	&= \int d\theta f^\text{ext}_x \psi - \gamma_\text{t} J_\text{n} + J_\text{m}\frac{d}{d x}(\frac{\gamma_\text{t} v_0}{D_\text{r}})\\
&- \frac{\gamma_\text{t}v_0}{k_\text{B}T} \int d\theta \sin\theta \left[(\Gamma_0 + \Gamma^\text{ext})\psi + \int d\mathbf{r}'\int d\theta' \Gamma \langle\hat{\psi}'\hat{\psi}\rangle\right].\nonumber
\end{align}
Notably, unlike the conventional force balance based on the intrinsic pressure in Eq.~(\ref{eq4}), Eq.~(\ref{eq13}) includes additional terms (i.e., the last two terms), arising from inhomogeneous activity or torques. In the presence of self-rotating or external torque, orientational interactions, or spatially inhomogeneous activity distributions, these terms resist reinterpretation as tensor divergences and cannot subsequently be absorbed into the active pressure. Such terms would complicate and even obscure the originally straightforward understanding to mechamical equilibrium. In contrast, as demonstrated by our theoretical derivations and simulations, the force balance based on the intrinsic pressure consistently remains its validity and simplicity across a broad range of active systems, showcasing its advantages.

\section{Conclusion}

In this study, we demonstrate that intrinsic pressure can be naturally chosen as the local pressure of dry active matter, and it provides a universal and convenient framework for the mechanical equilibrium of active systems. In this framework, the definition of the local pressure of dry active matter is consistent with the traditional, locally-determined form of pressure in passive systems, being a state function; while the self-propelling forces on the particles exerted by the implicit environment are treated as external body forces. The validity of this framework has been verified by performing simulations of diverse active systems. Our work thus recovers the conventional scenario of mechanical equilibrium in active systems, without invoking additional unfamiliar quantities.

\section{Acknowledgments}
This work was supported by the National Natural Science Foundation of China (Grants No.T2325027, No.12274448, No.T2350007, No.12404239 and No.12174041), National Key R\&D Program of China (2022YFF0503504) and the China Manned Space Engineering Program (KJZ-YY-NLT0502).

\bibliographystyle{apsrev4-2-titles}
\bibliography{myref.bib}

\end{document}